\documentclass[preprint]{aastex}

\begin{document}

\title{Physical Nature and Timing Variations of the Eclipsing System V407 Pegasi}
\author{Jae Woo Lee, Jang-Ho Park, Kyeongsoo Hong, Seung-Lee Kim, and Chung-Uk Lee}
\affil{Korea Astronomy and Space Science Institute, Daejeon 305-348, Korea}
\email{jwlee@kasi.re.kr, pooh107162@kasi.re.kr, kshong@kasi.re.kr, slkim@kasi.re.kr, leecu@kasi.re.kr}

\begin{abstract}
New multiband CCD photometry is presented for V407 Peg; the $R_{\rm C}$ light curves are the first ever compiled. 
Our light curves, displaying a flat bottom at secondary minimum and an O'Connell effect, were simultaneously analyzed 
with the radial-velocity (RV) curves given by Rucinski et al. (2008). The light changes of the system are best modeled 
by using both a hot spot on the secondary star and a third light. The model represents historical light curves also. 
All available minimum epochs, including our six timing measurements, have been examined and indicate that 
the eclipse timing variation is mainly caused by light asymmetries due to the spot activity detected in 
the light-curve synthesis. The hot spot may be produced as a result of the impact of the gas stream from the primary star. 
Our light and velocity solutions indicate that V407 Peg is a totally-eclipsing A-type overcontact binary with values of 
$q$=0.251, $i$=87$^\circ.6$, $\Delta T$=496 K, $f$=61 \%, and $l_3$=11$\sim$16 \%. Individual masses and radii of 
both components are determined to be $M_1$=1.72 M$_\odot$, $M_2$=0.43 M$_\odot$, $R_1$=2.15 R$_\odot$, and 
$R_2$=1.21 R$_\odot$. These results are very different from previous ones, which is probably caused by the light curves 
with distorted and inclined eclipses used in those other analyses. The facts that there are no objects optically related 
with the system and that the seasonal RVs show a large discrepancy in systemic velocity indicate that the third light source 
most likely arises from a tertiary component orbiting the eclipsing pair. 
\end{abstract}

{}

\section{INTRODUCTION}

V407 Peg ($\rm BD+14^{o} 5016$, 2MASS J23365535+1548063, TYC 1720-658-1; $V$=+9.28; F0V) was discovered to be an eclipsing 
variable by the Semi-Automatic Variability Search program at the Piwnice Observatory in Poland (Maciejewski et al. 2002, MKN). 
The light curves in the $BV$ bandpasses were typical of W UMa type and the maximum (Max II) following the secondary minimum 
was about 0.04 mag fainter than Max I. Maciejewski and collaborators (Maciejewski et al. 2003; Maciejewski \& Lig\c eza 2004) 
obtained double-line radial velocity (RV) curves with values of $K_1$=54.7 km s$^{-1}$, $K_2$=233.9 km s$^{-1}$, and 
$\gamma$=22.1 km s$^{-1}$. By combining the $BV$ light curves of MKN with the spectroscopic solutions, they reported that 
V407 Peg is an A-type overcontact binary with parameters of $i$=72$^\circ$.6, $\Delta$ ($T_{1}-T_{2}$)=329 K, $f$=54\%, 
$M_1$=1.48 M$_\odot$, and $M_2$=0.35 M$_\odot$. The light variations were explained by a hot spot located on the surface 
of the more massive primary star. 

Since then, Rucinski et al. (2008) measured 63 precise RVs at the David Dunlap Observatory (DDO) and derived spectroscopic 
elements ($K_1$=63.9 km s$^{-1}$, $K_2$=250.0 km s$^{-1}$, and $\gamma$=9.1 km s$^{-1}$) and a spectral type of F0V. 
These differ from previous results; the discrepancy is thought to be partially caused by the older RV curves, which lack 
coverage continuity. Recently, Deb \& Singh (2011) computed the binary parameters of V407 Peg using the spectroscopic solutions 
of Rucinski et al. (2008) and the $V$-band observations from the All Sky Automated Survey (ASAS) project (Pojmanski 1997, 2002).
Their results indicate that this system is a partially eclipsing binary with $i$=71$^\circ$.1, $\Delta$ ($T_{1}-T_{2}$)=862 K, 
$f$=81\%, $M_1$=1.92 M$_\odot$, and $M_2$=0.49 M$_\odot$. On the other hand, the orbital period of V407 Peg has been examined 
by Zasche (2011) only once. From the $O$--$C$ display based on newly derived minima and on historical data, he showed that 
the system has displaced secondary minima and suggested that the phenomenon is probably caused by asymmetrical light curves. 
Both eclipses from the MKN and ASAS data were asymmetric and distorted, but did not vary significantly with time.

Although V407 Peg has been studied photometrically and spectroscopically, the published light curves were somewhat incomplete 
or have been spread over several seasons. Intrinsic light variations for the system due to starspots and a third light source 
have not yet been considered in detail. Additionally, the eclipse timing variations still have not been described as 
conclusively as can be desired. In this paper, we present new multiband light curves and measure the physical properties of 
the eclipsing system from detailed studies of all available data, such as light and RV curves, and eclipse timings.

\section{CCD PHOTOMETRIC OBSERVATIONS}

New photometry of V407 Peg was performed on 14 nights from 2011 September 21 through December 27 in order to obtain 
multiband light curves. The observations were conducted using a PIXIS: 2048B CCD camera and a $BVR_{\rm C}$ filter set 
attached to the 61-cm reflector at Sobaeksan Optical Astronomy Observatory (SOAO) in Korea. The instrument and 
reduction method are the same as those described by Lee et al. (2013). Since the image field-of-view (FOV) was large 
enough to observe a few tens of nearby stars simultaneously, we monitored them along with the program target. 
Following the procedure described by Lee et al. (2010), potentially useful field stars were examined in detail 
for any peculiar light variations. Then, two non-variable candidates (TYC 1720-880-1 and TYC 1720-986-1) were combined 
using a weighted average to make an artificial reference star that would be optimal for our photometry.

A total of 4636 individual observations was obtained in the three bandpasses (1531 in $B$, 1568 in $V$, and 1537 in $R_{\rm C}$)
and a sample of those observations is provided in Table 1. The light curves of V407 Peg are plotted in Figure 1 as 
differential magnitudes between the variable and artificial comparison stars {\it versus} orbital phases; these data were 
computed according to the ephemeris for our hot-spot model on the less massive secondary star described in 
the following section. As shown in the figure, the SOAO observations are evenly distributed in phase and display 
the wavelength-dependent O'Connell effect for all bandpasses.

\section{LIGHT AND VELOCITY SOLUTIONS}

Our observations are typical of a short-period overcontact binary and display a flat bottom at secondary minimum, 
indicating that V407 Peg belongs to the A-type of W UMa stars. Further, as do the light curves of MKN and ASAS, 
the SOAO data are asymmetrical and Max I is brighter than Max II by about 0.074, 0.056, and 0.042 mag for the $B$, $V$, 
and $R_{\rm C}$ bandpasses, respectively. The light maxima (Max I and Max II) are displaced to around phases 0.238 and 0.755, 
respectively, and the primary minima are distorted and inclined, which may be caused by local photospheric inhomogeneities. 
In principle, these might be due to spot activity caused either by a magnetic dynamo or by impact from mass transfer 
between the components.

In order to obtain a unique set of the binary parameters for V407 Peg, our light curves were simultaneously modeled 
with the RV curves of Rucinski et al. (2008), taking into account proximity effects. We used contact mode 3 of 
the Wilson-Devinney synthesis code (Wilson \& Devinney 1971, WD) and a weighting scheme similar to that for 
the eclipsing systems RU UMi (Lee et al. 2008) and WZ Cyg (Lee et al. 2011). Table 2 lists the radial velocity and 
light-curve sets of the binary system simultaneously analyzed in this paper and their standard deviations ($\sigma$). 
The effective temperature of the larger and hotter primary star was assumed to be $T_{1}$=6980 K, according to 
its spectral type F0V given by Rucinski et al. (2008). The gravity-darkening exponents and the bolometric albedos were 
fixed at standard values of $g$=0.32 and $A$=0.5 for stars with common convective envelopes. The logarithmic bolometric 
($X$, $Y$) and monochromatic ($x$, $y$) limb-darkening coefficients were interpolated from the values of van Hamme (1993) 
and were used in concert with the model atmosphere option. Also, a synchronous rotation for both components and 
a circular orbit were adopted. In this paper, the subscripts 1 and 2 refer to the primary and secondary stars being 
eclipsed at Min I (at phase 0.0) and Min II, respectively.

Our analyses were carried out through three stages. In the first stage, the RV and light curves in Table 2 were solved, 
permitting no perturbations such as starspots or third light ($l_3$). The results are plotted as the short-dashed curves 
in Figure 1, for which the model light curves do not fit well the eclipse minima as well as light level at the first quadrature. 
In the second stage, the SOAO light curves were reanalyzed with the RV data by using the unperturbed solutions as 
the initial values and then considering a spot on either of the components. But, even the spotted solutions displayed as 
the long dashes in the same figure indicate that the observed eclipses are shallower than the computed ones and that 
the differences of the eclipse depths could be dependent on bandpasses. The fact may suggest a third light source in the system. 
Lastly, we modeled the observations including both spot and third light perturbations. Final results are given in Table 3, 
in which Model 1 and Model 2 represent the hot-spot model on the primary and secondary stars, respectively. 
Although it is not easy to discriminate between the two, we can see that Model 2 gives lightly smaller values for the sum 
of the residuals squared ($\Sigma W(O-C)^2$) than does Model 1. The synthetic light curves from Model 2 are plotted as 
the solid curves in Figure 1, while the synthetic RV curves are plotted in Figure 2 together with the measurements of 
Maciejewski \& Lig\c eza (2004) and Rucinski et al. (2008). As can be seen in the figures, our hot-spot model describes 
the SOAO multiband data quite well. Separate trials for a cool spot on either of the components were not so successful 
as the trials  for the hot-spot models. In all the procedures that have been described, we included an orbital eccentricity 
as a free parameter but found that the parameter retained a value of zero, which was within its margin of error.

To study the spot behavior of V407 Peg further and to examine whether our solutions can reasonably describe 
the historical light curves, we analyzed the MKN and ASAS data by adjusting only the orbital ephemeris ($T_0$ and $P$),
spot, and luminosity among the Model 2 parameters. These results appear in Table 4 and are plotted in Figure 3 
as the continuous curves. From the analyses, we can conclude that the hot-spot model on the secondary star satisfies 
all curves of V407 Peg quite well and gives a good representation of the binary system for both the photospheric and 
spot descriptions. As listed in Tables 3--4, the spot parameters have been almost constant with time since the discovery 
of the eclipsing pair and the stellar light ratios have not changed appreciably as a result of the spot modeling. 
Moreover, because V407 Peg should not have a deep common convective envelope as surmised from its spectral type and 
temperatures, the hot spot may be produced by stable mass transfer between the binary components.

\section{ECLIPSE TIMING VARIATIONS}

From the SOAO observations, eclipse timings and their errors for each filter were determined using the method of 
Kwee \& van Woerden (1956, KvW). Six new weighted times of minimum light are listed in the first column of Table 5, 
wherein two of the times (HJD 2,452,552.4416 and 2,452,582.3735) were derived by us from the individual measurements of MKN. 
In addition to these, 52 CCD timings have been collected from the literature (Maciejewski et al. 2002; Br\'at et al. 2011; 
Zasche 2011; Liakos \& Niarchos 2011; Gokay et al. 2012; Diethelm 2012, 2013, G\"ursoytrak et al. 2013). 

For ephemeris computations, weights for the eclipse timings were scaled as the inverse squares of their errors. 
In order to obtain a mean light ephemeris and to form an eclipse timing diagram, we applied a linear least-squares fit 
to all times of minimum light of V407 Peg and thus found an improved ephemeris, as follows:
\begin{equation}
 C = \mbox{HJD}~ 2452533.97494(21) + 0.636883596(47)E,
\end{equation}
where the parenthesized numbers are the 1$\sigma$-error values for the last digit of each term of the ephemeris. 
The resulting $O$--$C$ residuals calculated with equation (1) are plotted in Figure 4, in which the filled and 
open circles are the primary (Min I) and secondary (Min II) minima, respectively, from the literature. 
The up triangles represent the minimum times obtained using the KvW method from the MKN and SOAO data.

As displayed in Figure 4, the $O$--$C$ residuals from both eclipses are not in phase with each other and then 
this discrepancy can be attributed to the rotation of the apsidal line of the eccentric orbit due to tidal forces 
between the binary components. For apsidal motion, the $O$--$C$ residuals for primary minima are nearly 180$^\circ$ 
out of phase with those for the secondary. However, because the light-curve synthesis of V407 Peg indicates that 
the binary orbit is circular, the apsidal motion cannot explain the timing variations. Zasche (2011) suggested 
that the displaced secondary minima are caused by asymmetry of the light curves, resulting from a gas stream impact 
between the two components. In W UMa binaries, times of minimum light are shifted from the real conjunctions by 
asymmetrical eclipse minima due to stellar activity such as starspots (Kalimeris et al. 2002) and/or even by 
the method of measuring the eclipse timing (Maceroni \& van't Veer 1994). The light-curves synthesis method 
developed by WD is capable of extracting the conjunction instants and gives more and better information than that 
of the other methods, which do not consider spot activity and are based on observations during minimum alone. 
In reality, Lee et al. (2009) showed that the minimum epochs of the overcontact binary AR Boo have been systematically 
shifted by light asymmetries due to spot activity. 

In order to check for this possibility, we calculated the timings for each of all available eclipses using the WD code. 
The results are listed in the second column of Table 5 and are illustrated with the down triangles in Figure 4.
As shown in the third column of this table, there are systematic runs of differences between the KvW timings and 
the WD ones, which are negative for Min I and positive for Min II. These differences are caused by a hot spot 
on the secondary component presented to the observer after Min I and before Min II. In Figure 4, we can see that
the eclipse timing variations are mainly caused by the presence of a large hot spot which produces 
the asymmetrical light curves of V407 Peg.

\section{DISCUSSION AND CONCLUSIONS}

New multiband CCD light curves display complete eclipses leading to well-determined system parameters. Our light and 
velocity solutions indicate that V407 Peg is a totally-eclipsing A-type overcontact binary with a fill-out factor of 
about 60 \% and with a temperature difference of 496 K between the components. The binary model with a third light and 
a single hot spot on the less massive secondary component fits all available light curves quite well. We think that 
the hot spot may be produced as a result of the impact of the gas stream from the primary star, which is a dominant cause 
of the eclipse timing variations. The third light of 11$\sim$16 \% in all bandpasses could come from a tertiary object 
either gravitationally bound to or only optically related with the eclipsing pair. We do not see such an object 
optically resolved on our time-series CCD images. Although the third light source was not detected in the spectrograms 
of Rucinski et al. (2008), the two datasets of Maciejewski \& Lig\c eza (2004) and Rucinski et al. (2008), 
taken at the DDO and processed using the broadening function algorithm (Rucinski 2002), show a large discrepancy of 
about 13 km s$^{-1}$ in systemic velocity ($\gamma$). If the velocity difference is real, the third light most likely 
arises from the existence of a third component orbiting the eclipsing binary, which is a member of a triple system 
(Pribulla \& Rucinski 2006).

The simultaneous analysis of light and RV curves allowed us to compute the absolute parameters given in Table 6, 
together with those of Deb \& Singh (2011) for comparison. The luminosity ($L$) and bolometric magnitudes ($M_{\rm bol}$) 
were obtained by adopting $T_{\rm eff}$$_\odot$=5,780 K and $M_{\rm bol}$$_\odot$=+4.73 for solar values. In accordance 
with the unreliability in spectral classification, it was assumed that the temperature of each component had an error of 
200 K. The bolometric corrections (BCs) were obtained from the relation between $\log T_{\rm eff}$ and BC given by 
Torres (2010). With an apparent visual magnitude of $V$=+9.28 at maximum light, our computed light ratio, and 
the interstellar absorption of $A_{\rm V}$=0.186 (Schlafly \& Finkbeiner 2011), we determined the distance to the system 
to be 280$\pm$21 pc. The absolute parameters presented in this paper differ much from those of Deb \& Singh (2011). 
Considering our orbital inclination to be about 16 deg larger than that of earlier studies, the discrepancy is probably 
caused the ASAS light curves' displaying of inclined, partial eclipses. 

From our absolute parameters, it is possible to consider the evolutionary state of V407 Peg in terms of mass-radius, 
mass-luminosity, and the Hertzsprung-Russell (HR) diagrams. The locations of the component stars in these diagrams are shown 
in Figure 5, together with those of other well-studied overcontact binaries. The data are taken from the compilations of 
Yakut \& Eggleton (2005), Li et al. (2008), and Christopoulou et al. (2012). As can be seen in the figure, the binary components 
do conform to the general pattern of W UMa systems. The primary star lies in the main-sequence band, while the secondary is 
oversized and overluminous for its mass. This could be explained as a result of energy transfer from the more massive primary 
toward the less massive secondary star (e.g., Hilditch et al. 1988; Li et al. 2008). Because the light-curve synthesis of 
the system indicates that mass is moving between both components, a secular period variation may be produced through 
the mass transfer in the system. More systematic and continuous high-resolution observations (photometry and spectroscopy) 
will help to identify and understand the orbital period variation of V407 Peg.

\acknowledgments{ }

We would like to thank the staff of the Sobaeksan Optical Astronomy Observatory for assistance during our observations. 
This research has made use of the Simbad database maintained at CDS, Strasbourg, France.  This work was supported by 
the KASI (Korea Astronomy and Space Science Institute) grant 2014-1-400-06.

\clearpage
\begin{figure}
 \includegraphics[]{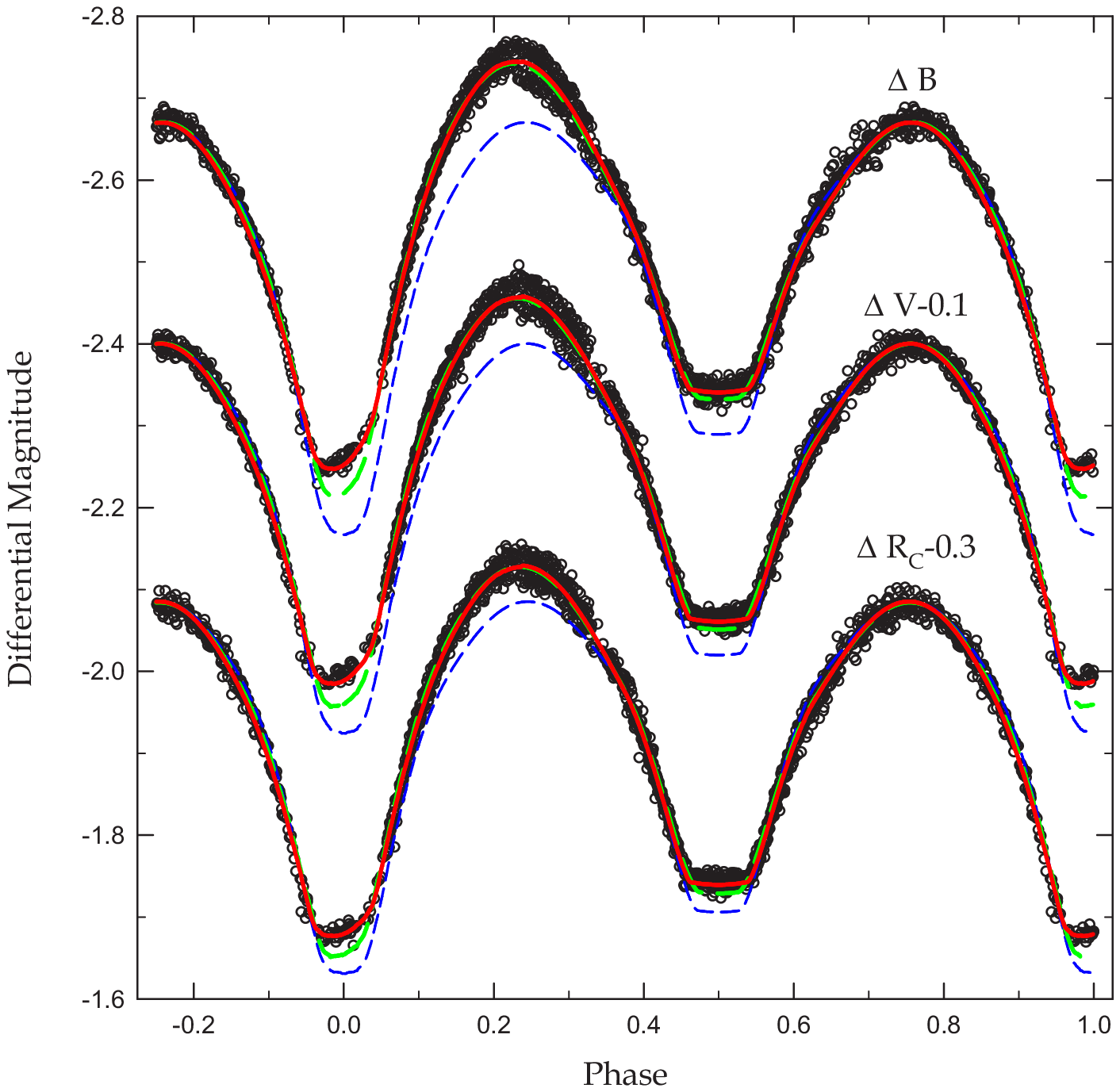}
 \caption{$BVR_{\rm C}$ observations of V407 Peg with fitted model light curves. The short and long dashes are computed without
 and with a spot, respectively, and the solids represent the solutions obtained with both a spot and a third light ($l_3$).
 See the text for details. }
 \label{Fig1}
\end{figure}

\begin{figure}
 \includegraphics[]{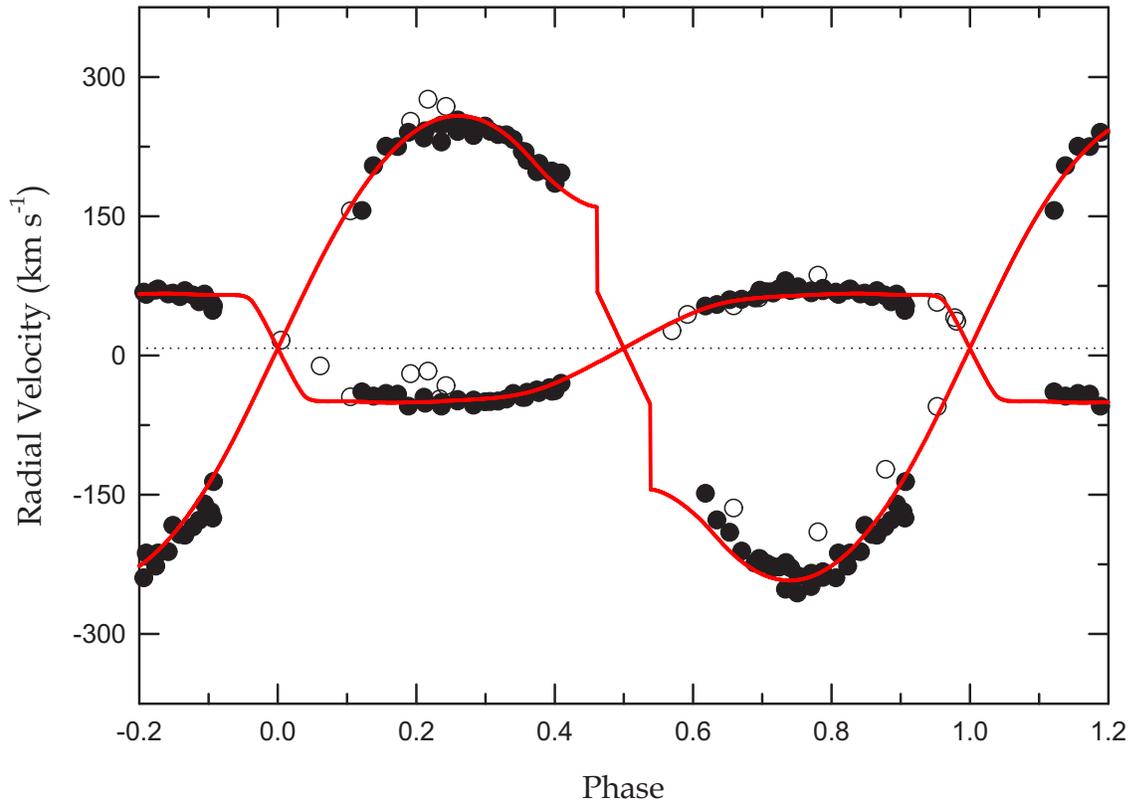}
 \caption{Radial-velocity curves of V407 Peg. The open and filled circles are the measures of Maciejewski \& Lig\c eza (2004)
 and Rucinski et al. (2008), respectively, while the solid curves denote the results from consistent light and RV curve analysis. 
 The dotted line refers to the system velocity of 7.9 km s$^{-1}$.}
\label{Fig2}
\end{figure}

\begin{figure}
 \includegraphics[]{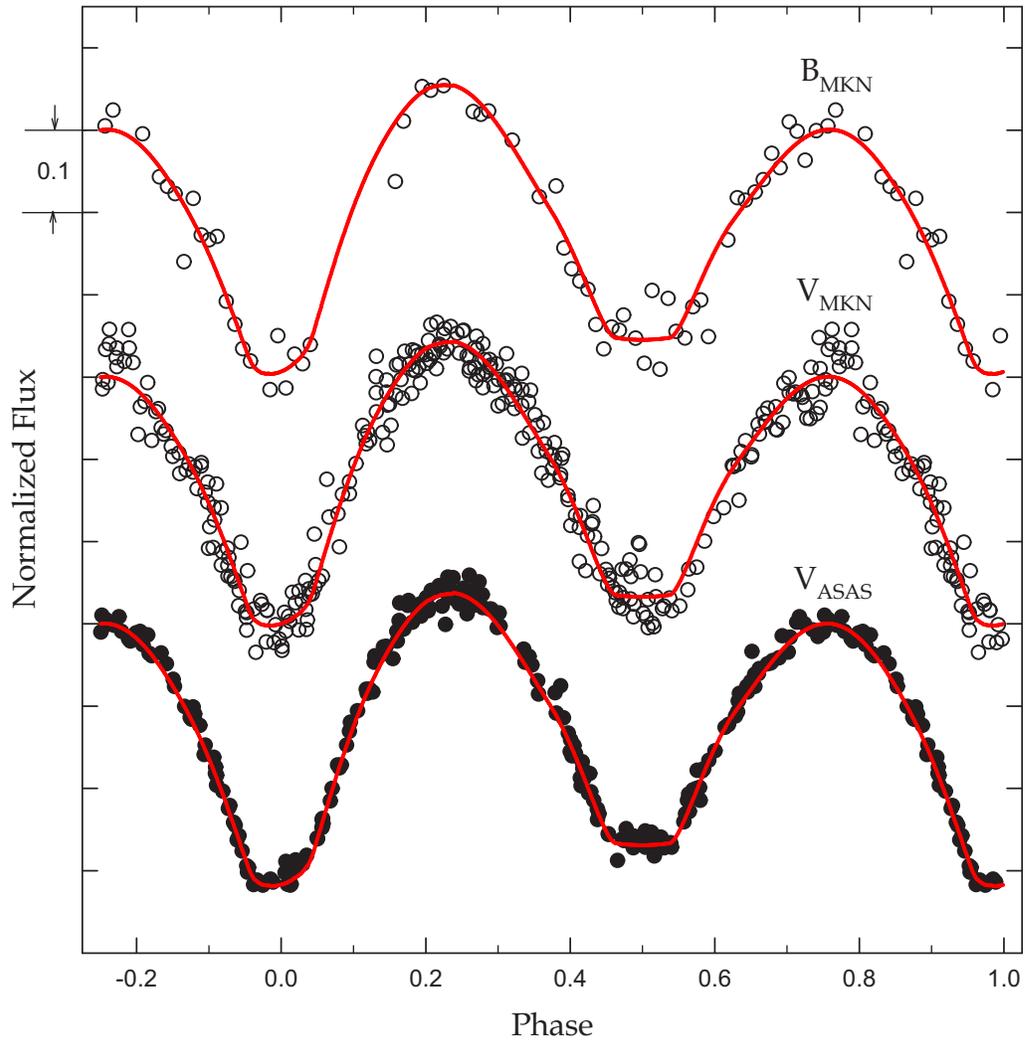}
 \caption{Historical light curves of V407 Peg. The open and filled circles are the measures of MKN and ASAS, respectively. 
  The continuous curves represent solutions obtained with the spot-model parameters listed in Table 4.}
 \label{Fig3}
\end{figure}

\begin{figure}
 \includegraphics[]{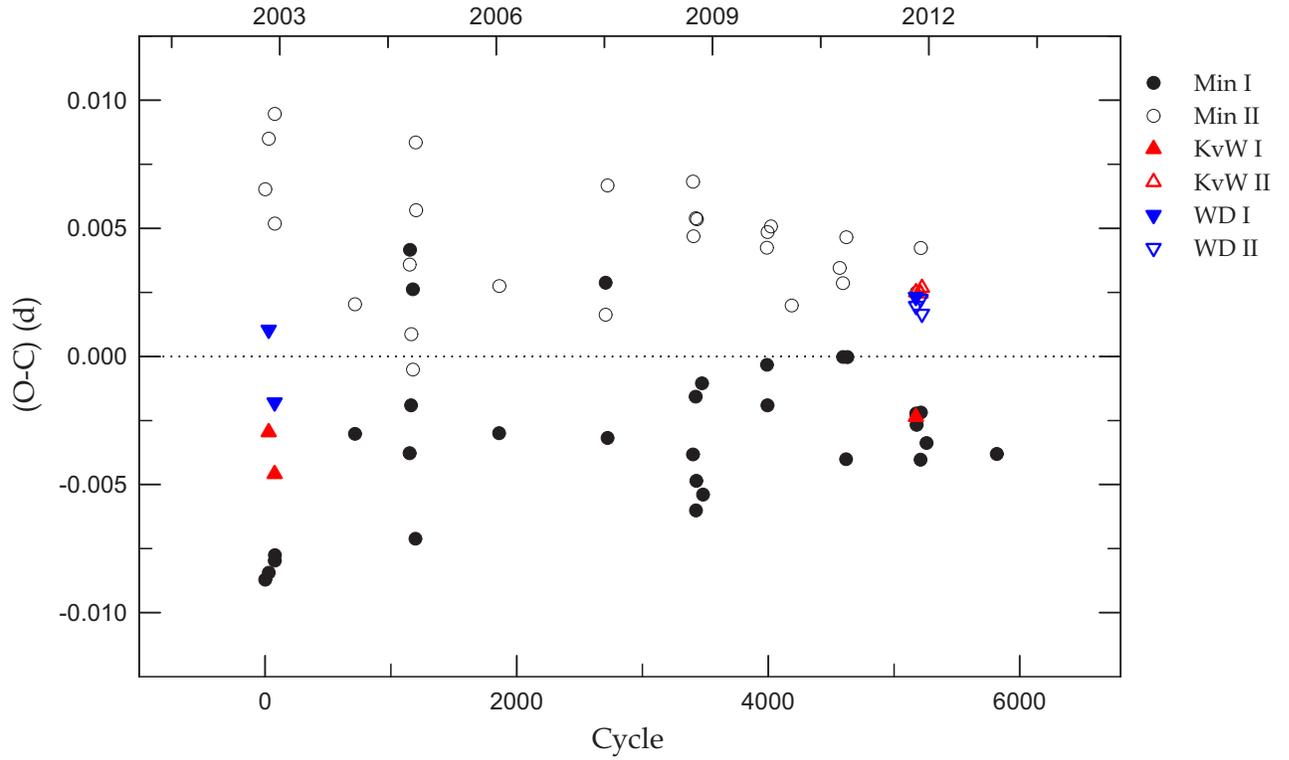}
 \caption{$O$--$C$ diagram of V407 Peg. The filled and open circles are the primary (Min I) and secondary (Min II) minima, 
 respectively, from the literature. The triangle symbols represent the minimum times in Table 5, obtained by us from 
 the MKN and SOAO data using the KvW method and the WD synthesis code. }
 \label{Fig4}
\end{figure}

\begin{figure}
 \includegraphics[]{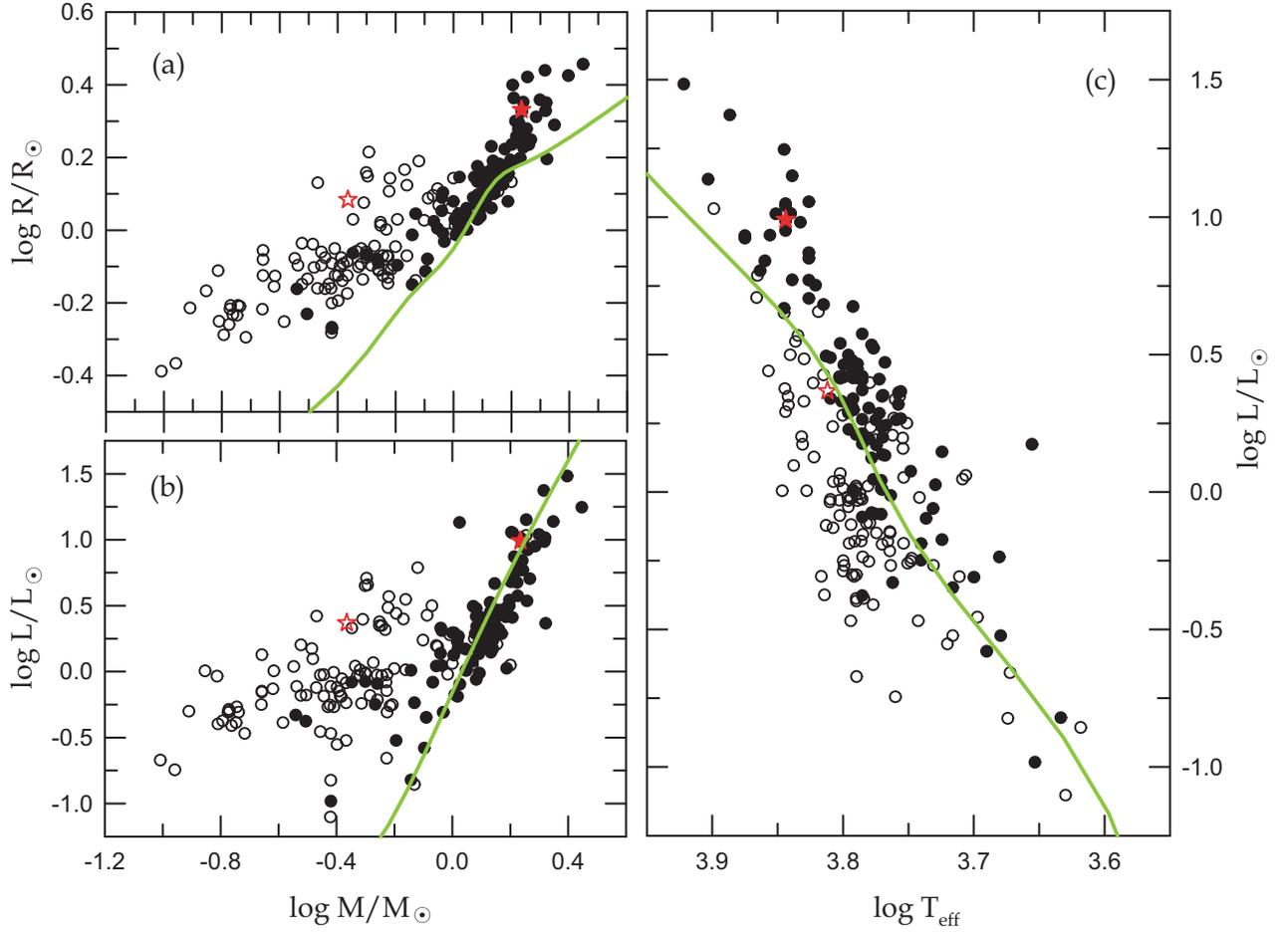}
 \caption{Plots of (a) mass-radius, (b) mass-luminosity, and (c) HR diagrams for W UMa-type eclipsing systems. The filled 
 and  open circles represent the more massive and less massive components, respectively. The star symbols denote the locations 
 of the components of V407 Peg in these diagrams. The solid lines correspond to the zero-age main sequence stars calculated 
 as having a solar metallicity of $Z$=0.02 by Tout et al. (1996). }
 \label{Fig5}
\end{figure}

\clearpage
\begin{deluxetable}{crcrcr}
\tabletypesize{\small}
\tablewidth{0pt} 
\tablecaption{CCD photometric observations of V407 Peg.}
\tablehead{
\colhead{HJD} & \colhead{$\Delta B$} & \colhead{HJD} & \colhead{$\Delta V$} & \colhead{HJD} & \colhead{$\Delta R_{\rm C}$} 
}
\startdata
2,455,826.06512 & $-$2.3954  &  2,455,826.06669 & $-$2.2111  &  2,455,826.06751 &  $-$2.0998   \\
2,455,826.06896 & $-$2.4170  &  2,455,826.06942 & $-$2.2368  &  2,455,826.06794 &  $-$2.1066   \\
2,455,826.07045 & $-$2.4292  &  2,455,826.09630 & $-$2.3818  &  2,455,826.06826 &  $-$2.1152   \\
2,455,826.07079 & $-$2.4268  &  2,455,826.09763 & $-$2.3923  &  2,455,826.06977 &  $-$2.1183   \\
2,455,826.07121 & $-$2.4312  &  2,455,826.09959 & $-$2.3939  &  2,455,826.09666 &  $-$2.2579   \\
2,455,826.07157 & $-$2.4383  &  2,455,826.10086 & $-$2.4051  &  2,455,826.09814 &  $-$2.2544   \\
2,455,826.07192 & $-$2.4703  &  2,455,826.10224 & $-$2.4118  &  2,455,826.09857 &  $-$2.2703   \\
2,455,826.07888 & $-$2.4889  &  2,455,826.10352 & $-$2.4134  &  2,455,826.09995 &  $-$2.2629   \\
2,455,826.07922 & $-$2.4898  &  2,455,826.10471 & $-$2.4174  &  2,455,826.10123 &  $-$2.2749   \\
2,455,826.08321 & $-$2.5064  &  2,455,826.10596 & $-$2.4188  &  2,455,826.10382 &  $-$2.2875   \\
\enddata
\tablecomments{This table is available in its entirety in machine-readable and Virtual Observatory (VO) forms 
in the online journal. A portion is shown here for guidance regarding its form and content.}
\end{deluxetable}

\begin{deluxetable}{lccc}
\tablewidth{0pt}
\tablecaption{Radial velocity and light-curve sets for V407 Peg.}
\tablehead{
\colhead{Reference}          & \colhead{Season} & \colhead{Data type} & \colhead{$\sigma$} }
\startdata
Rucinski et al. (2008)       & 2006$-$2007      & RV1                 & 5.5 km s$^{-1}$    \\
                             &                  & RV2                 & 15.2 km s$^{-1}$   \\
SOAO                         & 2011             & $B$                 & 0.011 mag          \\
                             &                  & $V$                 & 0.010 mag          \\
                             &                  & $R_{\rm C}$         & 0.010 mag          \\
\enddata
\end{deluxetable}

\begin{deluxetable}{lccccc}
\tabletypesize{\small}  
\tablewidth{0pt}
\tablecaption{RV and light curve parameters of V407 Peg.}
\tablehead{
\colhead{Parameter}                     & \multicolumn{2}{c}{Model 1}              && \multicolumn{2}{c}{Model 2}              \\ [1.0mm] \cline{2-3} \cline{5-6} \\[-2.0ex]
                                        & \colhead{Primary} & \colhead{Secondary}  && \colhead{Primary} & \colhead{Secondary}
}
\startdata                                                                         
$T_0$ (HJD)                             & \multicolumn{2}{c}{2,452,534.6105(29)}   && \multicolumn{2}{c}{2,452,534.6100(28)}   \\
$P$ (day)                               & \multicolumn{2}{c}{0.63688436(57)}       && \multicolumn{2}{c}{0.63688436(53)}       \\
$\gamma$ (km s$^{-1}$)                  & \multicolumn{2}{c}{7.9(7)}               && \multicolumn{2}{c}{7.9(7)}               \\
$a$ (R$_\odot$)                         & \multicolumn{2}{c}{4.02(3)}              && \multicolumn{2}{c}{4.02(3)}              \\
$q$                                     & \multicolumn{2}{c}{0.251(3)}             && \multicolumn{2}{c}{0.251(3)}             \\
$i$ (deg)                               & \multicolumn{2}{c}{87.6(6)}              && \multicolumn{2}{c}{87.6(5)}              \\
$T$ (K)                                 & 6980              & 6535(6)              && 6980              & 6484(5)              \\
$\Omega$                                & 2.275(2)          & 2.275                && 2.261(3)          & 2.261                \\
$\Omega_{\rm in}$                       & \multicolumn{2}{c}{2.356}                && \multicolumn{2}{c}{2.356}                \\
$f$ (\%)                                & \multicolumn{2}{c}{51}                   && \multicolumn{2}{c}{61}                   \\ 
$X$, $Y$                                & 0.639, 0.254      & 0.638, 0.242         && 0.639, 0.254      & 0.638, 0.241         \\
$x_{B}$, $y_{B}$                        & 0.790, 0.275      & 0.804, 0.233         && 0.790, 0.275      & 0.806, 0.228         \\
$x_{V}$, $y_{V}$                        & 0.691, 0.289      & 0.708, 0.275         && 0.691, 0.289      & 0.711, 0.274         \\
$x_{R_{\rm C}}$, $y_{R_{\rm C}}$        & 0.594, 0.293      & 0.614, 0.285         && 0.594, 0.293      & 0.617, 0.284         \\
$l$/($l_{1}$+$l_{2}$+$l_{3}$){$_{B}$}   & 0.700(3)          &  0.149               && 0.716(4)          &  0.147               \\
$l$/($l_{1}$+$l_{2}$+$l_{3}$){$_{V}$}   & 0.712(3)          &  0.164               && 0.717(4)          &  0.162               \\
$l$/($l_{1}$+$l_{2}$+$l_{3}$){$_{R_{\rm C}}$}  & 0.719(3)   &  0.176               && 0.718(4)          &  0.173               \\
{\it $l_{3B}$$\rm ^a$}                  & \multicolumn{2}{c}{0.151(3)}             && \multicolumn{2}{c}{0.137(4)}             \\
{\it $l_{3V}$$\rm ^a$}                  & \multicolumn{2}{c}{0.124(3)}             && \multicolumn{2}{c}{0.121(4)}             \\
{\it $l_{3R_{\rm C}}$$\rm ^a$}          & \multicolumn{2}{c}{0.105(3)}             && \multicolumn{2}{c}{0.109(4)}             \\
$r$ (pole)                              & 0.4880(4)         & 0.2689(10)           && 0.4913(7)         & 0.2727(15)           \\
$r$ (side)                              & 0.5332(7)         & 0.2830(13)           && 0.5381(10)        & 0.2877(19)           \\
$r$ (back)                              & 0.5637(9)         & 0.3379(33)           && 0.5700(14)        & 0.3492(53)           \\
$r$ (volume)$\rm ^b$                    & 0.5295            & 0.2967               && 0.5344            & 0.3023               \\ [1.0mm]
\multicolumn{6}{l}{Spot parameters:}                                                                                           \\ 
Colatitude (deg)                        & 88.0(2)           & \dots                && \dots             & 86.0(4)              \\
Longitude (deg)                         & 301.1(2)          & \dots                && \dots             & 130.9(2)             \\
Radius (deg)                            & 21.4(1)           & \dots                && \dots             & 47.3(2)              \\
$T$$\rm _{spot}$/$T$$\rm _{local}$      & 1.109(1)          & \dots                && \dots             & 1.131(1)             \\
$\Sigma W(O-C)^2$                       & \multicolumn{2}{c}{0.0106}               && \multicolumn{2}{c}{0.0104}               \\ 
\enddata
\tablenotetext{a}{Value at 0.75 phase.}
\tablenotetext{b}{Mean volume radius.}
\end{deluxetable}

\begin{deluxetable}{lcc}
\tablewidth{0pt}
\tablecaption{Spot and luminosity parameters for historical light curves.}
\tablehead{
\colhead{Parameter}                       &  \colhead{MKN}      &  \colhead{ASAS}               
}                                                                                                                                                
\startdata                                                                                                                                       
$T_0$ (HJD)$\rm ^a$                       &  2,552.44592(70)    &  3,889.90252(28)    \\
$P$ (day)                                 &  0.636882(20)       &  0.63688428(24)     \\
Colatitude$_2$ (deg)                      &  85.6(7.4)          &  85.1(4.1)          \\
Longitude$_2$ (deg)                       &  146.4(3.4)         &  140.0(1.5)         \\
Radius$_2$ (deg)                          &  49.1(3.1)          &  45.6(8)            \\
$T$$\rm _{spot,2}$/$T$$\rm _{local,2}$    &  1.131(6)           &  1.115(3)           \\
$l_1$/($l_{1}$+$l_{2}$+$l_{3}$){$_{B}$}   &  0.696(45)          &  \dots              \\
$l_1$/($l_{1}$+$l_{2}$+$l_{3}$){$_{V}$}   &  0.713(40)          &  0.718(30)          \\
{\it $l_{3B}$$\rm ^b$}                    &  0.161(42)          &  \dots              \\
{\it $l_{3V}$$\rm ^b$}                    &  0.126(37)          &  0.119(20)          \\
\enddata
\tablenotetext{a}{HJD 2,450,000 is suppressed.}
\tablenotetext{b}{Value at 0.75 phase.}
\end{deluxetable}

\begin{deluxetable}{cccccl}
\tablewidth{0pt}
\tablecaption{New minimum timings of V407 Peg.}
\tablehead{
KvW$\rm^{a}$        &  WD$\rm^{a}$            &  Difference$\rm^{b}$  &  Filter  &  Min  &  References
}
\startdata
2,552.4416$\pm$0.0011~~~  &  2,552.4456$\pm$0.0014~~~  &  $-$0.0040~~  &  $V$            &  I   &  MKN                \\
2,582.3735$\pm$0.0020~~~  &  2,582.3763$\pm$0.0015~~~  &  $-$0.0028~~  &  $BV$           &  I   &  MKN                \\
5,829.20831$\pm$0.00035   &  5,829.21299$\pm$0.00012   &  $-$0.00468   &  $BVR_{\rm C}$  &  I   &  This paper         \\
5,830.16850$\pm$0.00027   &  5,830.16796$\pm$0.00013   &  $+$0.00054   &  $BVR_{\rm C}$  &  II  &  This paper         \\
5,853.09626$\pm$0.00019   &  5,853.09605$\pm$0.00013   &  $+$0.00021   &  $BVR_{\rm C}$  &  II  &  This paper         \\
5,860.10220$\pm$0.00016   &  5,860.10119$\pm$0.00008   &  $+$0.00101   &  $BVR_{\rm C}$  &  II  &  This paper         \\
\enddata
\tablenotetext{a}{HJD 2,450,000 is suppressed. }
\tablenotetext{b}{Differences between columns (1) and (2). }
\end{deluxetable}

\begin{deluxetable}{lccccc}
\tablewidth{0pt} 
\tablecaption{Physical properties of V407 Peg.}
\tablehead{
\colhead{Parameter}       & \multicolumn{2}{c}{Deb \& Singh (2011)}  && \multicolumn{2}{c}{This Work}               \\ [1.0mm] \cline{2-3} \cline{5-6} \\[-2.0ex]
                          & \colhead{Primary} & \colhead{Secondary}  && \colhead{Primary} & \colhead{Secondary}            
}
\startdata
$a$ (R$_\odot$)           & \multicolumn{2}{c}{4.182$\pm$0.027}      && \multicolumn{2}{c}{4.019$\pm$0.029}         \\          
$q$                       & \multicolumn{2}{c}{0.256$\pm$0.006}      && \multicolumn{2}{c}{0.2515$\pm$0.0032}       \\          
$i$ ($^\circ$)            & \multicolumn{2}{c}{71.11$\pm$0.67}       && \multicolumn{2}{c}{87.55$\pm$0.54}          \\          
$T$ (K)                   & 7300$\pm$184      & 6438$\pm$141         && 6980$\pm$200       & 6484$\pm$200           \\
$\Omega$                  & 2.240$\pm$0.009   & 2.240                && 2.2611$\pm$0.0026  & 2.2611                 \\          
$r$                       & 0.470$\pm$0.003   & 0.312$\pm$0.003      && 0.5344$\pm$0.0010  & 0.3023$\pm$0.0025      \\ [0.5mm]
$M$ (M$_\odot$)           & 1.922$\pm$0.023   & 0.492$\pm$0.015      && 1.718$\pm$0.026    & 0.432$\pm$0.009        \\
$R$ (R$_\odot$)           & 1.965$\pm$0.018   & 1.305$\pm$0.015      && 2.146$\pm$0.015    & 1.214$\pm$0.013        \\
$\log$ $g$ (cgs)          & \dots             & \dots                && 4.010$\pm$0.009    & 3.905$\pm$0.013        \\
$\rho$ (g cm$^3)$         & \dots             & \dots                && 0.245$\pm$0.006    & 0.341$\pm$0.013        \\
$L$ (L$_\odot$)           & 9.83$\pm$1.17     & 2.62$\pm$0.29        && 9.80$\pm$1.13      & 2.33$\pm$0.29          \\
$M_{\rm bol}$ (mag)       & \dots             & \dots                && $+$2.25$\pm$0.13   & $+$3.81$\pm$0.14       \\
BC (mag)                  & \dots             & \dots                && $+$0.03$\pm$0.01   & $+$0.01$\pm$0.01       \\
$M_{\rm V}$ (mag)         & \dots             & \dots                && $+$2.22$\pm$0.13   & $+$3.80$\pm$0.14       \\
Distance (pc)             & \dots             & \dots                && \multicolumn{2}{c}{280$\pm$21}              \\
\enddata
\end{deluxetable}


\newpage
\begin{thebibliography}{}
\bibitem[Brat et al(2011)]{brat2011} Br\'at, L., et al. 2011, Open Eur. J. Var. Stars, 137, 1
\bibitem[Christopoulou et al(2012)]{christopoulou2012} Christopoulou, P.-E., Papageorgiou, A., Vasileiadis, T., \& Tsantilas, S. 2012, AJ, 144, 149
\bibitem[Deb \& Singh(2011)]{deb2011} Deb, S., \& Singh, H. P. 2011, MNRAS, 412, 1787
\bibitem[Diethelm(2012)]{diethelm2012} Diethelm, R. 2012, Inf. Bull. Variable Stars, 6011, 1
\bibitem[Diethelm(2013)]{diethelm2013} Diethelm, R. 2013, Inf. Bull. Variable Stars, 6042, 1
\bibitem[Gokay et al(2012))]{gokay2012} Gokay, G., et al. 2012, Inf. Bull. Variable Stars, 6039, 1
\bibitem[Gursoytrak et al(2013))]{gursoytrak2013} G\"ursoytrak, H., et al. 2013, Inf. Bull. Variable Stars, 6075, 1
\bibitem[Hilditch et al(1988)]{hilditch1988} Hilditch, R. W., King, D. J., \& McFarlane, T. M. 1988, MNRAS, 231, 341
\bibitem[Kalimeris et al(2002)]{kalimeris2002} Kalimeris, A., Rovithis-Livaniou, H., \& Rovithis, P. 2002, A\&A, 387, 969
\bibitem[Kwee \& van Woerden(1956)]{kwee1956} Kwee, K. K., \& van Woerden, H. 1956, Bull. Astron. Inst. Netherlands, 12, 327
\bibitem[Lee et al(2008)]{lee2008} Lee, J. W., Kim, C.-H., Kim, S.-L., Lee, C.-U., Han, W., \& Koch, R. H. 2008, PASP, 120, 720
\bibitem[Lee et al(2009)]{lee2009} Lee, J. W., Youn, J.-H., Lee, C.-U., Kim, S.-L., \& Koch, R. H. 2009, AJ, 138, 478
\bibitem[Lee et al(2010)]{lee2010} Lee, J. W., et al. 2010, AJ, 139, 898
\bibitem[Lee et al(2011)]{lee2011} Lee, J. W., Kim, S.-L., Lee, C.-U., Kim, H.-I., Park, J.-H., \& Hinse, T. C. 2011, AJ, 142, 12
\bibitem[Lee et al(2013)]{lee2010} Lee, J. W., Hinse, T. C., \& Park, J.-H. 2013, AJ, 145, 100
\bibitem[Li et al(2008)]{li2008} Li, L., Zhang, F. Han, Z., Jiang, D., \& Jiang, T. 2008, MNRAS, 387, 97
\bibitem[Liakos \& Niarchos(2012)]{liakos2012} Liakos, A., \& Niarchos, P. 2011, Inf. Bull. Variable Stars, 6005, 1
\bibitem[Maceroni \& van't Veer(1994)]{maceroni1994} Maceroni, C., \& van't Veer, F. 1994, A\&A, 289, 871
\bibitem[Maciejewski et al(2002)]{maciejewski2002} Maciejewski, G., Karska, A., \& Niedzielski, A. 2002, Inf. Bull. Variable Stars, 5343, 1 (MKN)
\bibitem[Maciejewski et al(2003)]{maciejewski2003} Maciejewski, G., Lig\c eza, P., \& Karska, A. 2003, Inf. Bull. Variable Stars, 5400, 1
\bibitem[Maciejewski \& Lig\c eza(2004)]{maciejewski2004} Maciejewski, G., \& Lig\c eza, P. 2004, Inf. Bull. Variable Stars, 5504, 1
\bibitem[Pojmanski(1997)]{pojmanski1997} Pojmanski, G. 1997, Acta Astron., 47, 467
\bibitem[Pojmanski(2002)]{pojmanski2002} Pojmanski, G. 2002, Acta Astron., 52, 397
\bibitem[Pribulla \& Rucinski(2006)]{pribulla2006} Pribulla, T., \& Rucinski, S. M. 2006, AJ, 131, 2986
\bibitem[Rucinski(2002)]{rucinski2002} Rucinski, S. M. 2002, AJ, 124, 1746
\bibitem[Rucinski et al(2008)]{rucinski2008} Rucinski, S. M., et al. 2008, AJ, 136, 586
\bibitem[Schlafly \& Finkbeiner(2011)]{schlafly2011} Schlafly, E. F., \& Finkbeiner, D. P. 2011, AJ, 737, 103
\bibitem[Torres(2010)]{torres2010} Torres, G. 2010, AJ, 140, 1158
\bibitem[Tout(1996)]{tout1996} Tout, C. A., Pols, O. R., Eggleton, P. P., \& Han, Z. 1996, MNRAS, 281, 257
\bibitem[Wilson \& Devinney(1971)]{wilson1971} Wilson, R. E., \& Devinney, E. J. 1971, ApJ, 166, 605
\bibitem[Yakut \& Eggleton(2005)]{yakut2005} Yakut, K., \& Eggleton, P. P. 2005, ApJ, 629, 1055
\bibitem[Van Hamme(1993)]{van1993} Van Hamme, W. 1993, AJ, 106, 209
\bibitem[Zasche(2011)]{zasche2011} Zasche, P. 2011, Inf. Bull. Variable Stars, 5991, 1
\end{thebibliography}
\end{document}